\DocumentMetadata{
	lang		= eng-US, 	
	pdfstandard	= A-3u,		
	pdfversion	= 1.7, 		
}

\documentclass[colorlinks,upint,subscriptcorrection,varvw,hyphenate,balance,greek,russian,vietnamese,german]{asmeconf} 
\usepackage[main=USenglish]{babel}

\usepackage{stfloats}

\hypersetup{%
	pdfauthor={John H. Lienhard},									  
	pdftitle={ASME Conference Paper LaTeX Template},                  
	pdfkeywords={ASME conference paper, LaTeX template, BibTeX style},
	pdfsubject = {Describes the asmeconf LaTeX template},			  
}

\allowdisplaybreaks 

\begin{document}


\ConfName{Proceedings of the ASME 2026\linebreak Fluids Engineering Division Summer Meeting}
\ConfAcronym{FEDSM2026}
\ConfDate{July 26-29, 2026} 
\ConfCity{Bellevue, Washington}
\PaperNo{FEDSM2025-185195}

%

\title{AI-Accelerated Operator Learning Framework for Rarefied Microflows} 
 
%
%
%

\SetAuthors{%
   Ehsan Roohi\affil{1}\CorrespondingAuthor{} 
	}

\SetAffiliation{1}{Department of Mechanical and Industrial Engineering, University of Massachusetts Amherst, Amherst, MA 01003, USA}


\maketitle

\versionfootnote{Documentation for \texttt{asmeconf.cls}: Version~\versionno, \today.}




\begin{abstract}
The high computational cost of kinetic solvers such as DSMC remains a major challenge in rarefied flow simulations. This work presents a unified framework combining deep neural networks and neural operators to accelerate kinetic and hybrid solvers while preserving physical fidelity. GPU-native DNN surrogates eliminate costly moment-closure operations in Fokker–Planck methods, achieving significant speedups without accuracy loss, while physics-guided and shock-aware DeepONet architectures enable accurate, data-efficient modeling of multi-regime micro-nozzle, micro-step, and hypersonic flows. Extensions including ensemble uncertainty quantification and family-of-experts strategies further enhance robustness across wide Mach and Knudsen number ranges. Together, these results demonstrate a scalable and physics-consistent pathway toward real-time surrogate modeling in rarefied gas dynamics..
\end{abstract}



\section{Introduction}
High-fidelity simulation of rarefied gas dynamics is a cornerstone of modern aerospace and
micro-scale engineering, underpinning applications ranging from hypersonic re-entry and
high-altitude flight to micro-propulsion systems, MEMS devices, and vacuum technologies.
In these regimes, the molecular mean free path becomes comparable to the characteristic
length scale of the flow, leading to strong non-equilibrium effects such as velocity slip,
temperature jump, Knudsen layers, and breakdown of local thermodynamic equilibrium.
As a result, classical continuum-based models, including the Navier--Stokes--Fourier
equations, fail to provide physically reliable predictions, particularly in the slip and
transitional regimes~\cite{roohi2025}.

To accurately resolve such flows, kinetic methods rooted in the Boltzmann equation have
become the accepted gold standard. Among these, the Direct Simulation Monte Carlo (DSMC)
method remains the most widely used tool due to its robustness and physical fidelity~\cite{bird1994}.
However, DSMC suffers from an inherently high computational cost, as it requires explicit
tracking of molecular motion and stochastic collisions while resolving the smallest spatial
and temporal scales dictated by the mean free path and collision time. This cost becomes
especially prohibitive for near-continuum regimes, shock-containing flows, and many-query
tasks such as parametric sweeps, uncertainty quantification, inverse design, and optimization.
Consequently, the computational burden of kinetic solvers has emerged as a critical
bottleneck in the practical design and analysis of rarefied flow systems~\cite{bird2013}.

Hybrid kinetic approaches, such as particle-based Fokker--Planck (FP) methods~\cite{jenny2010}, offer a
partial remedy by replacing discrete collision processes with deterministic drift--diffusion
models in velocity space. While these formulations substantially reduce stochastic noise and
relax timestep constraints, they introduce new computational stiffness through expensive
moment-closure operations, which must be solved repeatedly and locally throughout the
domain~\cite{gorji2011,gorji2014,gorji2015}. Thus, despite algorithmic advances, the core challenge persists: achieving
DSMC-level accuracy at a computational cost compatible with real-time or iterative design
workflows.

In parallel with advances in kinetic modeling, recent years have witnessed rapid growth in
machine-learning-based surrogate modeling for complex physical systems~\cite{peyvan2024,peyvan2026}. Deep neural
networks (DNNs) have demonstrated remarkable expressive power for approximating
high-dimensional nonlinear mappings, offering the promise of orders-of-magnitude speedups
once trained~\cite{roohi2026}. Early data-driven surrogates applied to rarefied flows successfully reproduced
DSMC results for specific configurations, but often required large training datasets and
exhibited limited robustness, physical consistency, and extrapolation capability. These
limitations highlighted the need to embed physical structure, inductive bias, and domain
knowledge directly into learning frameworks.

Physics-informed learning marked a major conceptual shift by incorporating governing
equations, constraints, or invariants into the training process~\cite{raissi2019}. However, conventional
Physics-Informed Neural Networks (PINNs) are designed to approximate a single solution
instance. They are therefore ill-suited for parametric studies, where predictions are required
across continuous ranges of operating conditions, geometries, or rarefaction levels. Neural
operator frameworks, such as the Deep Operator Network (DeepONet)~\cite{peyvan2024}, address this
limitation by learning the solution operator itself, i.e., the mapping from parameters to full
solution fields. This operator-learning paradigm is particularly well aligned with rarefied
flow problems, which naturally involve strong parametric dependence on Knudsen number, Mach number, pressure ratio, and geometry.

Despite their promise, direct application of neural operators to rarefied gas dynamics
introduces several nontrivial challenges. These include handling multi-regime behavior
within a single flow, resolving sharp non-equilibrium structures such as shocks and
recirculation zones, maintaining physical consistency under sparse training data, and
integrating surrogates efficiently into existing high-performance simulation pipelines.
Addressing these challenges requires more than generic architectures; it demands
problem-specific strategies that respect the underlying kinetic physics.

Motivated by these considerations, this work synthesizes a series of complementary
methodologies that collectively establish a robust framework for accelerating kinetic solvers
and constructing physics-consistent surrogates for rarefied gas dynamics. The presented
approaches span both \emph{solver-level acceleration} and \emph{operator-level surrogate
modeling}. At the solver level, GPU-native deep neural networks are employed to replace
computationally expensive deterministic components of particle-based FP methods, entirely
eliminating CPU--GPU communication overhead and achieving near-theoretical speedups.
At the operator level, physics-guided and shock-aware DeepONet architectures are developed
to model complex rarefied flows, including micro-step separation, shock-dominated
micro-nozzle flows, hypersonic cylinder aerodynamics, and lid-driven cavity flows across
wide Knudsen number ranges.

A unifying theme of this work is the deliberate integration of physical insight into machine
learning design choices. This includes physics-guided feature spaces aligned with shock
locations, zonal and curriculum-based loss functions that prioritize non-equilibrium regions,
ensemble-based uncertainty quantification for extrapolative regimes, and modular
expert-based strategies for handling wide parametric variability. Together, these techniques
enable data-efficient learning from expensive DSMC datasets while preserving fidelity to the
underlying physics.

By consolidating these advances into a single methodological narrative, the present work
demonstrates that deep learning and neural operators are not merely post-processing tools
for kinetic simulations, but can serve as principled, scalable, and physically grounded
components of next-generation rarefied flow solvers. The resulting framework provides a
clear pathway toward real-time surrogate modeling, rapid design exploration, and
multi-query analysis in regimes where traditional kinetic methods remain computationally
prohibitive.

\section{Fokker--Planck Solver Acceleration via a GPU-Native DNN Closure}

Particle-based Fokker--Planck (FP) methods provide an attractive alternative to DSMC in low-to-moderate Knudsen
regimes by replacing stochastic collisions with a deterministic drift--diffusion process in velocity space. However,
advanced FP formulations (e.g., cubic-FP) introduce a severe computational bottleneck through the \emph{moment-closure}
step: at every timestep and in every cell, high-order moments must be computed from particles and a dense linear system
(e.g., $9\times 9$ for cubic-FP) must be solved to recover closure coefficients for stress and heat flux. As highlighted in
our study, this closure step can dominate the runtime and limits the scalability of FP solvers on modern GPUs
\cite{gorji2011,gorji2015}. In this work, we replace the expensive closure solve with a deep neural network (DNN)
surrogate deployed \emph{entirely on the GPU}, thereby removing the dominant cost while preserving physical fidelity.

\subsection{Algorithmic Overview}
The proposed acceleration follows a four-phase workflow (data generation $\rightarrow$ offline training $\rightarrow$ parameter
extraction $\rightarrow$ GPU-native deployment). In the baseline physics loop, particles are moved and boundary conditions
are applied, full moments (including high-order moments) are gathered, then linear systems are assembled and solved to
obtain the closure coefficients before particle velocities are evolved. We target the linear-system closure as the primary
bottleneck \cite{gorji2011}. In the accelerated loop, the closure solve is replaced by a single GPU-resident forward pass of
a trained MLP: (i) a one-time parameter extraction dumps all network weights/biases and scaling statistics into an \texttt{.npz}
file, (ii) a \texttt{LITE} moment routine computes only the 16 low-order input features required by the model (skipping the
expensive high-order moment calculations used by the full physics solver), and (iii) the forward pass is implemented using
pure CuPy matrix operations to eliminate CPU--GPU I/O overhead. The surrogate predicts the 9 closure
coefficients (stress/heat-flux parameters) directly from the local feature vector, enabling a drop-in replacement for the
closure step in the time-marching loop.

\subsection{Micro-Cavity Results}
\paragraph{Result 1: Strong-scaling speedup reaches the Amdahl limit.}
For a 2D lid-driven cavity benchmark (5000 timesteps, $\sim$630k particles), the physics-based solver spends $\approx$41--42\%
of total runtime in the closure solve. Replacing this step with the GPU-native surrogate yields a measured speedup of
$1.71\times$ on a $25\times 25$ grid and $1.73\times$ on a $100\times 100$ grid (Table~\ref{tab:fp_speedup}). A strong-scaling
analysis shows that the \emph{maximum} achievable speedup is $\approx 1.727\times$ based on Amdahl's law, and the observed
$1.73\times$ indicates the surrogate closure is effectively ``negligible-cost'' relative to the remaining particle operations
\cite{gorji2011}. This demonstrates a fundamental bottleneck shift: after acceleration, the dominant cost becomes particle
moment gathering rather than closure solving.

\begin{table}[t]
\centering
\caption{Strong-scaling performance for 2D cavity: replacing the closure solve with the GPU-native DNN.}
\label{tab:fp_speedup}
\begin{tabular}{lcccc}
\hline
Grid & $T_{\text{Physics}}$ & $T_{\text{ML}}$ & $T_{\text{Solver}}$ & Speedup \\
\hline
$25\times 25$ & 84.96 s & 49.82 s & 35.14 s & $1.71\times$ \\
$100\times 100$ & 81.72 s & 47.35 s & 34.37 s & $1.73\times$ \\
\hline
\end{tabular}
\end{table}

\begin{figure*}[t]
\centering

\begin{subfigure}[b]{\columnwidth}
\centering{
  \includegraphics[width=0.85\linewidth]{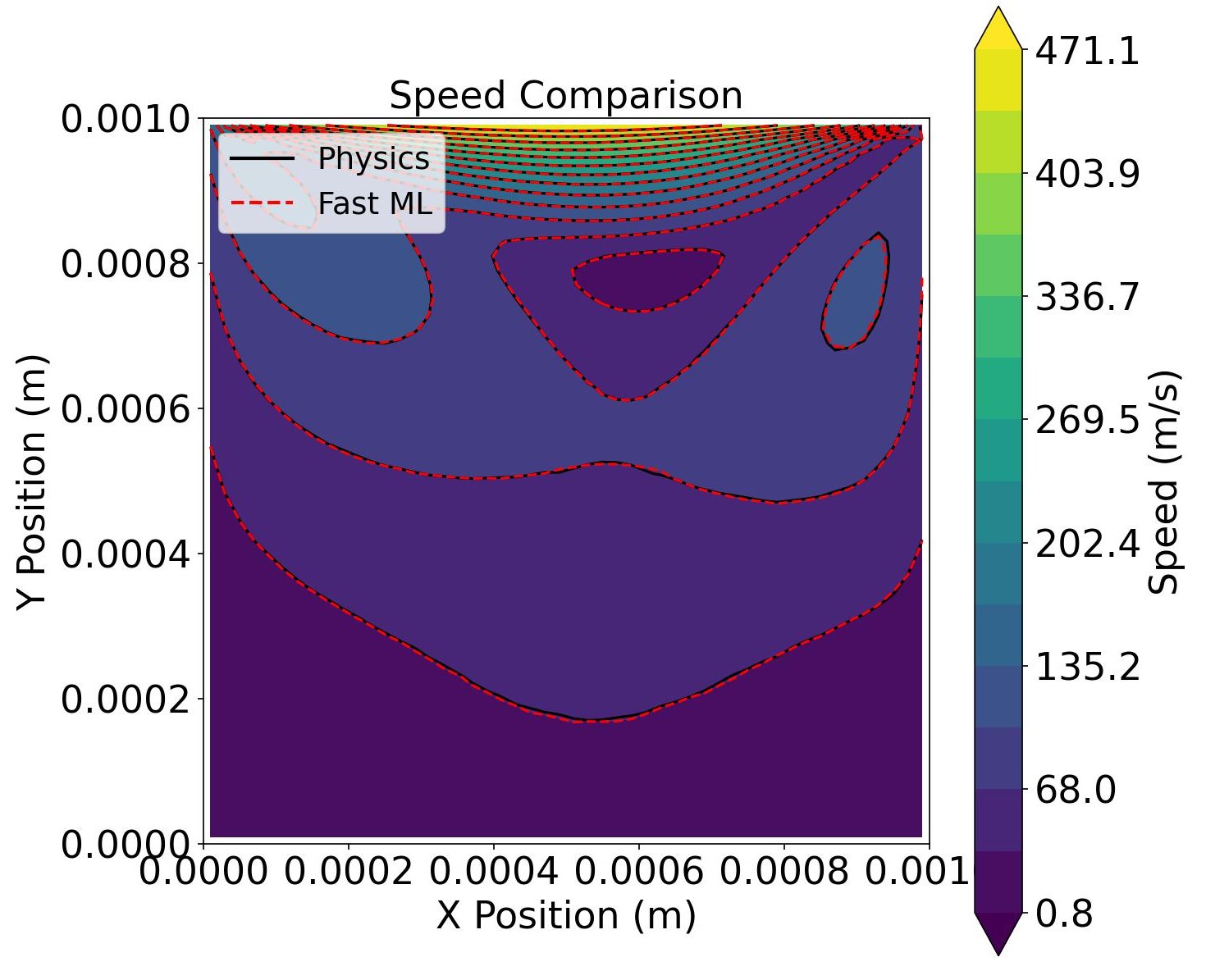}
}
\subcaption{Speed comparison\label{fig:speed_800ms}}
\end{subfigure}%
\hspace*{\columnsep}%
\begin{subfigure}[b]{\columnwidth}
\centering{
  \includegraphics[width=0.85\linewidth]{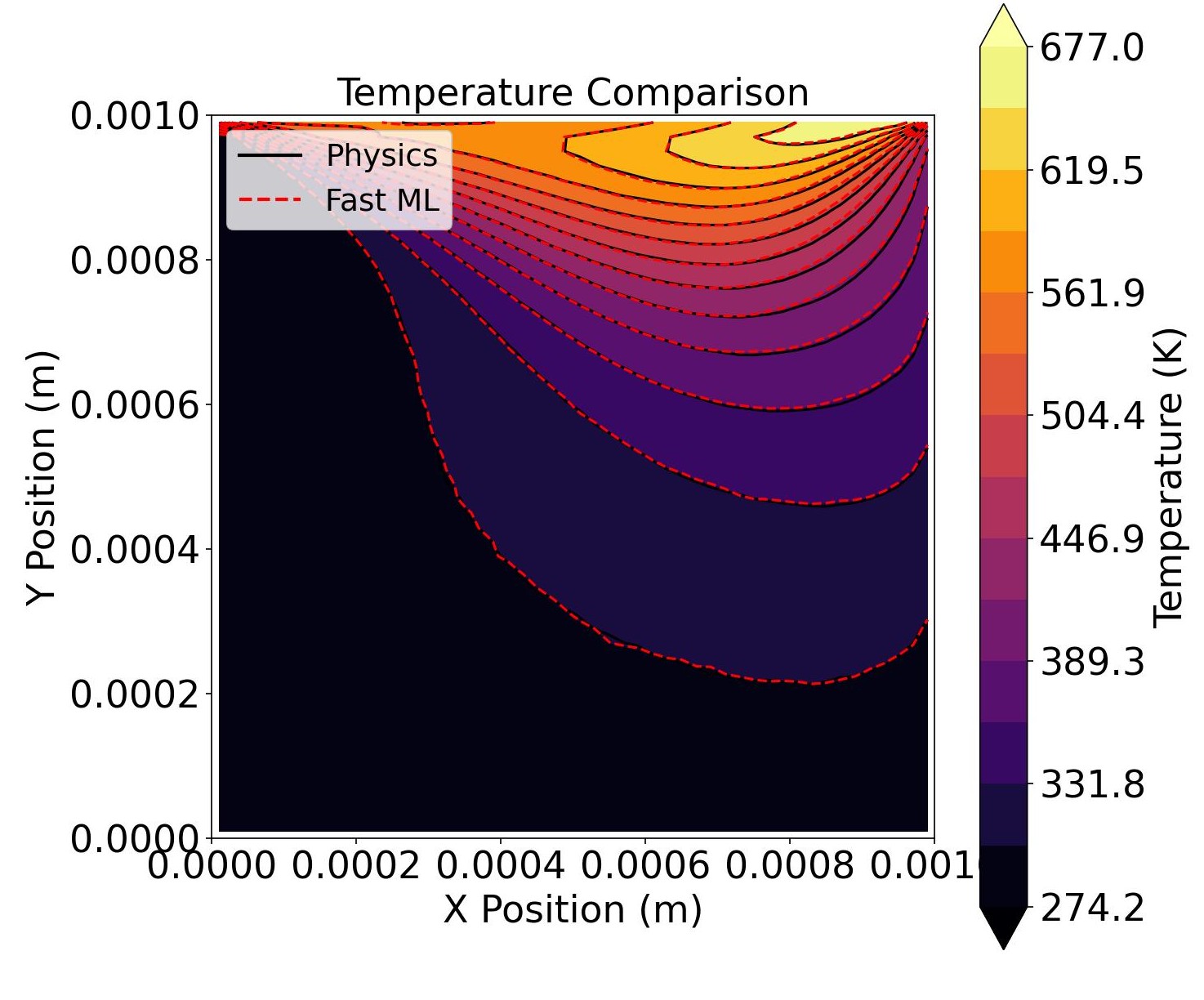}
}
\subcaption{Temperature comparison\label{fig:temp_800ms}}
\end{subfigure}

\vspace{0.35cm}

\begin{subfigure}[b]{\columnwidth}
\centering{
  \includegraphics[width=0.85\linewidth]{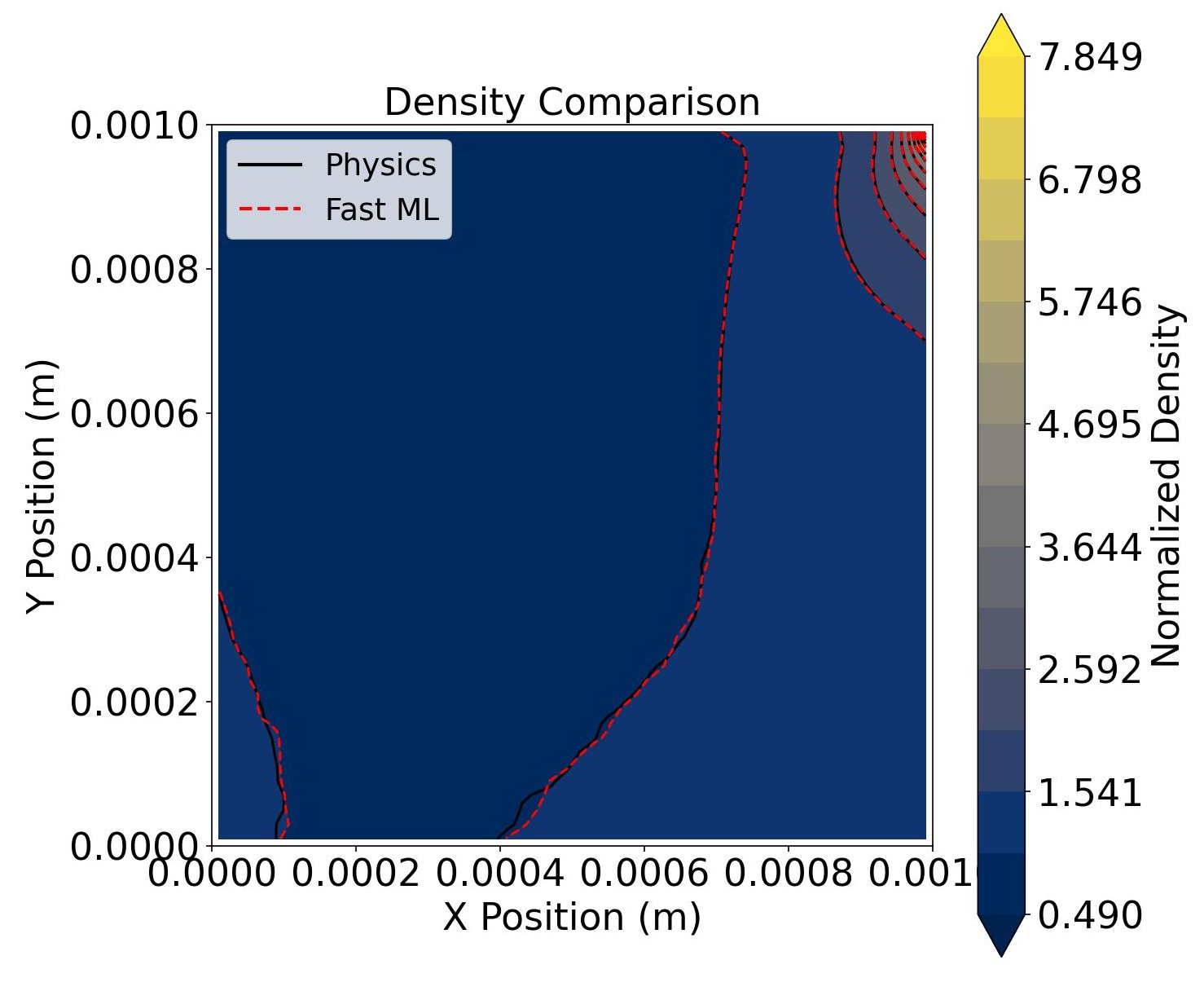}
}
\subcaption{Density comparison\label{fig:density_800ms}}
\end{subfigure}

\caption{
\textbf{Extrapolation test at $U_{\mathrm{lid}}=800~\mathrm{m/s}$.}
Comparison of 2D contour fields between the full \textbf{Physics} solver (solid black lines)
and the \textbf{Fast ML} solver (dashed red lines).
The surrogate model is trained only on lower-velocity cases ($U_{\mathrm{lid}}\le 200~\mathrm{m/s}$),
yet maintains excellent agreement at four times the maximum training velocity, demonstrating robust generalization.
\label{fig:generalization_test_800ms}}
\end{figure*}

\paragraph{Result 2: Robust extrapolation to a hyper-velocity cavity case.}
To test generalization, a robust cavity surrogate was trained on a lid-velocity sweep
$U_{\text{lid}}=\{50,100,200,400,600\}\,$m/s. The trained model was then deployed \emph{without any retraining}
to predict an extreme $U_{\text{lid}}=800\,$m/s case (a $4\times$ extrapolation beyond the maximum training velocity).
Despite the significantly stronger compressibility and viscous-heating effects (peak temperature exceeding $\sim$670 K), the surrogate produces near-perfect agreement with the physics solver: see Fig.~\ref {fig:generalization_test_800ms}. It captures the lid shear layer, and the hot-spot location and contour topology in the top-right region. Error analysis
shows that deviations are not random; they remain largely localized to the sharpest-gradient corner region, providing a clear
pathway for targeted physics-informed corrections or localized hybridization when needed.


\section{Neural-Operator Surrogate for Hypersonic Rarefied Flow over a Cylinder}

Hypersonic flow over blunt bodies represents a canonical and highly challenging problem in rarefied
gas dynamics, combining strong shock waves, steep gradients in thermodynamic variables, and
significant non-equilibrium effects. Such configurations are central to atmospheric re-entry vehicles
and high-speed aerospace systems, where accurate prediction of surface and near-field flow properties
is critical for aerodynamic heating and force estimation. In the rarefied regime, these flows cannot be
reliably captured by continuum-based solvers, and high-fidelity kinetic methods such as DSMC are
required. However, the extreme computational cost of DSMC severely limits its use in parametric
studies across wide Mach-number ranges.

\begin{figure*}[t]
\centering
\includegraphics[width=0.65\textwidth]{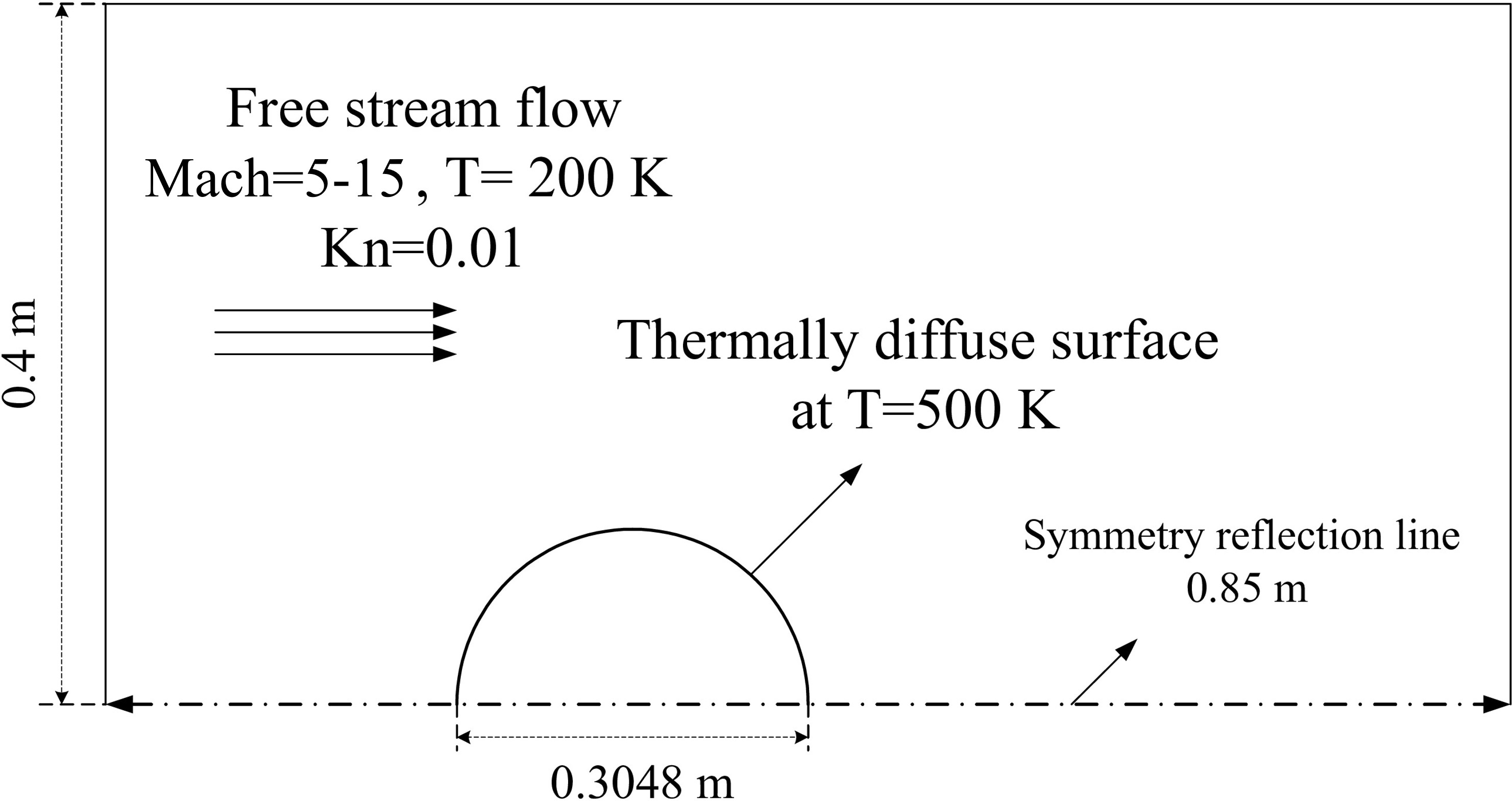}
\caption{
\textbf{Hypersonic cylinder configuration and boundary conditions.}
Schematic of the computational setup for rarefied hypersonic flow over a circular cylinder.
The freestream conditions are $M_\infty=5$--$15$, $T_\infty=200~\mathrm{K}$, and $Kn=0.01$.
The cylinder surface is modeled as a thermally diffuse wall at $T_w=500~\mathrm{K}$.
A symmetry line is applied along the centerline to reduce the computational domain.
}
\label{fig:cylinder_schematic}
\end{figure*}

In this work, we employ a data-driven Deep Operator Network (DeepONet) ensemble to construct
an efficient surrogate model for two-dimensional hypersonic flow over a circular cylinder in the
rarefied regime. 
Figure~\ref{fig:cylinder_schematic} summarizes the computational domain and boundary conditions
for the rarefied hypersonic cylinder case considered in this study.
The surrogate is trained on high-fidelity DSMC datasets spanning Mach numbers
from $M=5$ to $M=14$, and is designed to learn the nonlinear operator mapping the freestream Mach
number to the full spatial distribution of flow-field variables. The DeepONet architecture consists of
a branch network encoding the parametric dependence on Mach number and a trunk network encoding
the spatial coordinates, whose outputs are combined to reconstruct the flow field.

To enhance robustness and provide principled uncertainty quantification, a deep ensemble strategy
is adopted, where multiple DeepONet models are trained independently with different initializations.
The ensemble mean provides the final prediction, while the ensemble variance serves as a measure of
epistemic uncertainty, which is particularly important for extrapolative predictions in hypersonic
regimes.

The trained surrogate demonstrates excellent generalization capability across both interpolation and
extrapolation tasks. For interpolation, the model accurately predicts unseen Mach numbers
($M=7, 9, 12,$ and $14$) with near-perfect agreement against DSMC reference solutions.
More importantly, in the extrapolation case, the ensemble successfully predicts the full flow field at $M=15$, despite being trained exclusively on data up to $M=14$. 

\begin{figure*}[t]
\centering

\begin{subfigure}[b]{\columnwidth}
\centering{
  \includegraphics[width=0.85\linewidth]{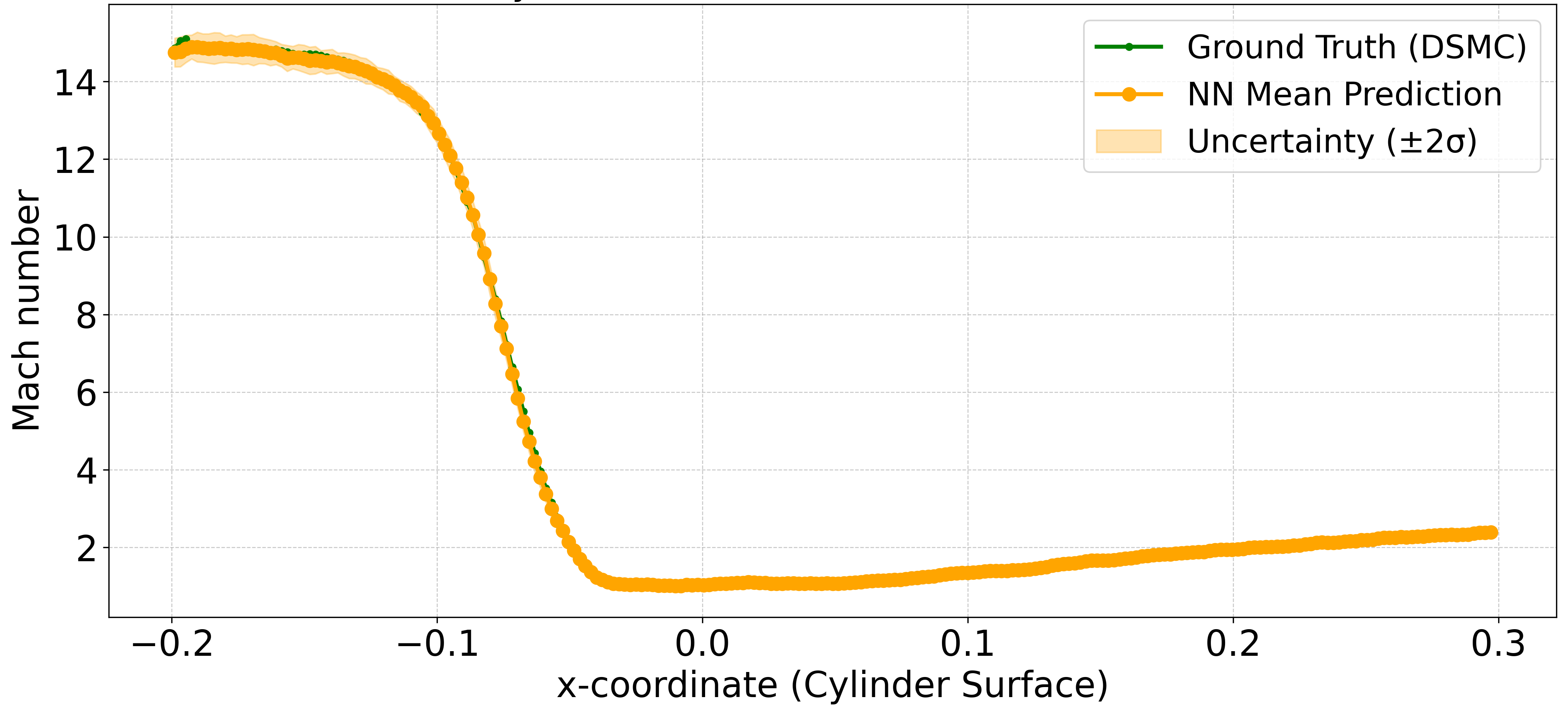}
}
\subcaption{Surface Mach number distribution\label{fig:cyl_Ma}}
\end{subfigure}%
\hspace*{\columnsep}%
\begin{subfigure}[b]{\columnwidth}
\centering{
  \includegraphics[width=0.85\linewidth]{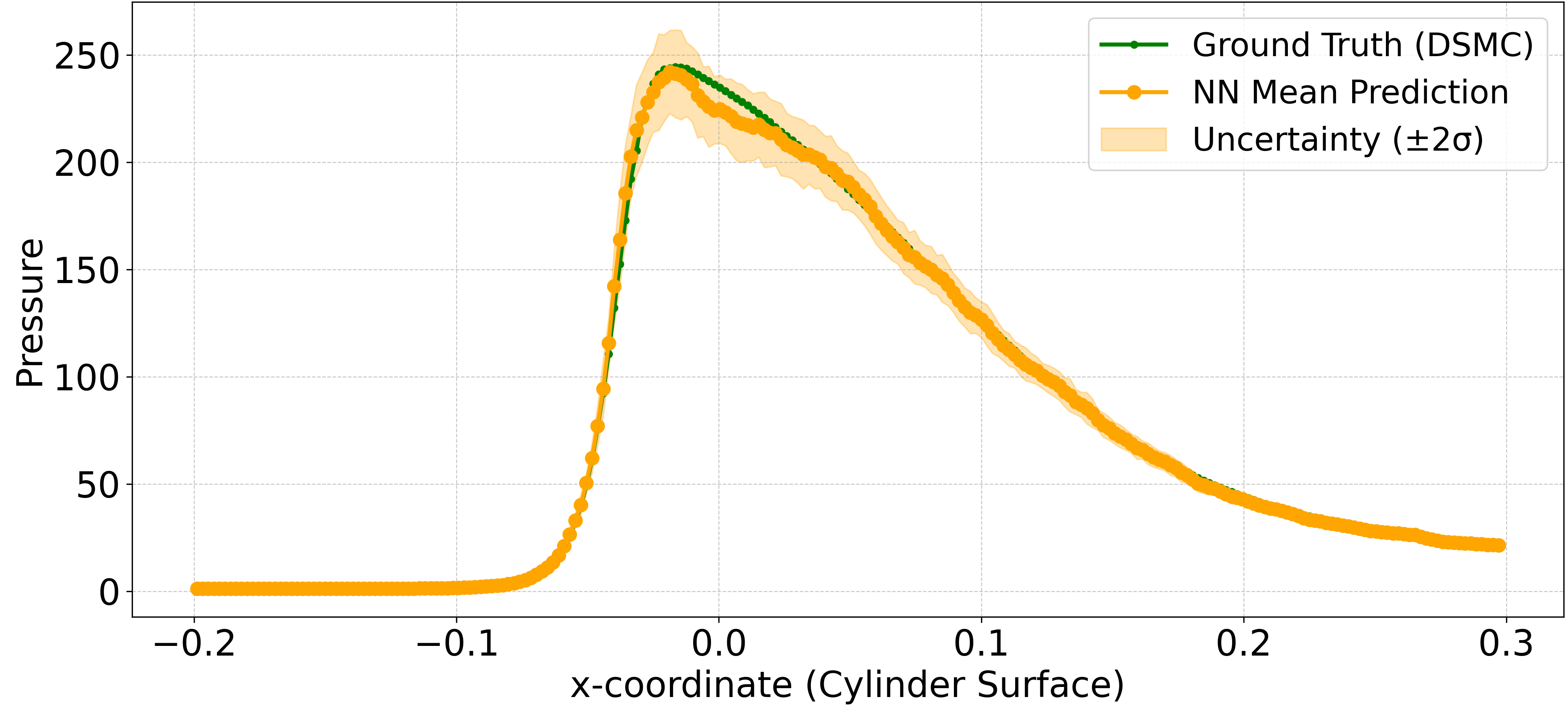}
}
\subcaption{Surface pressure distribution\label{fig:cyl_P}}
\end{subfigure}

\vspace{0.35cm}

\begin{subfigure}[b]{\columnwidth}
\centering{
  \includegraphics[width=0.85\linewidth]{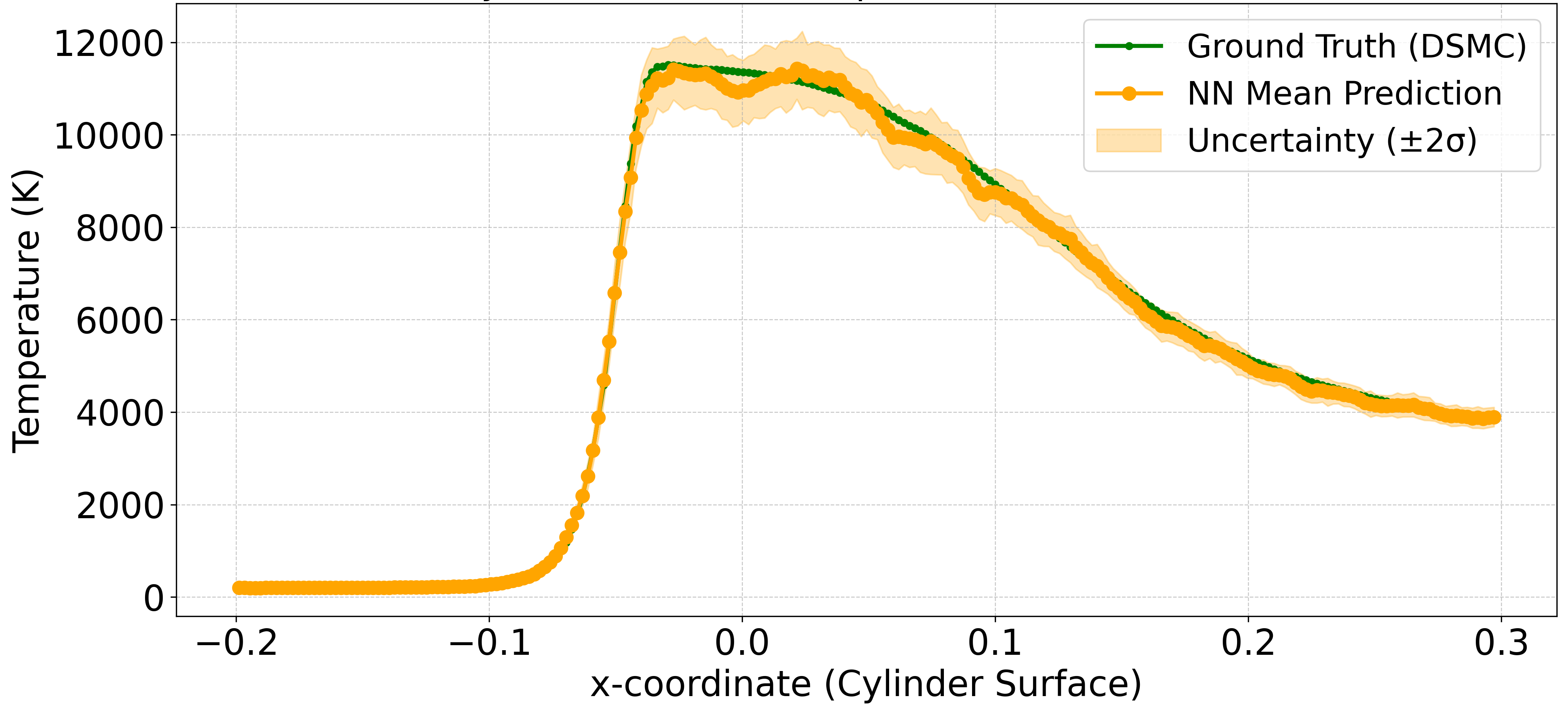}
}
\subcaption{Surface temperature distribution\label{fig:cyl_T}}
\end{subfigure}

\caption{
\textbf{Surface property distributions over a cylinder at $M=15$.}
Comparison between DSMC reference data (green symbols) and DeepONet ensemble predictions
(solid orange lines), with shaded regions indicating the epistemic uncertainty
($\pm 2\sigma$).
The surrogate accurately captures shock-induced deceleration, pressure amplification,
and extreme thermal loading, while maintaining narrow uncertainty bounds except
in the immediate shock region.
}
\label{fig:cylinder_surface_profiles}
\end{figure*}

Figure~\ref{fig:cylinder_surface_profiles} presents surface distributions of Mach number,
pressure, and temperature along the cylinder for the extrapolation case at $M=15$.
The DeepONet ensemble prediction (solid orange line) is compared against the DSMC
reference solution (green symbols), with shaded bands indicating the epistemic uncertainty
estimated from the ensemble ($\pm 2\sigma$).
The Mach-number profile accurately captures the sharp deceleration across the detached
bow shock, followed by the gradual recovery in the downstream region.
Similarly, the pressure distribution exhibits excellent agreement in both the shock-induced
compression peak and the subsequent relaxation along the cylinder surface.
The temperature profile reproduces the intense post-shock heating, reaching peak values
above $10{,}000~\mathrm{K}$, and correctly follows the downstream thermal relaxation trend.
Notably, the uncertainty bands remain narrow across most of the surface and widen only
slightly in the immediate shock region, reflecting increased model uncertainty in zones
of extreme gradients.
Overall, these results demonstrate that the neural-operator surrogate not only preserves
high-fidelity kinetic physics under mild extrapolation, but also provides physically
meaningful confidence bounds for hypersonic rarefied flow predictions.

Overall, this cylinder benchmark demonstrates that neural-operator-based surrogates can serve as
powerful and reliable tools for rapid prediction and parametric exploration of rarefied hypersonic
flows, bridging the gap between DSMC-level fidelity and real-time design requirements.


\subsection{DeepONet surrogate for rarefied backward-facing step: Knudsen- and geometry-parametric learning}
The backward-facing step in rarefied gas dynamics provides a stringent benchmark for data-driven surrogates because the flow topology changes nonlinearly with the Knudsen number. As rarefaction increases, separation and reattachment weaken, the primary recirculation bubble shrinks, and in highly rarefied regimes the corner vortex can nearly disappear, yielding markedly different streamline patterns and velocity gradients. This regime-dependent evolution implies that the mapping from the control parameter to the solution field can exhibit sharp transitions, making the operator approximation problem challenging for conventional smooth-interpolation networks.

To address this difficulty, a Deep Operator Network (DeepONet) surrogate is employed to learn the operator that maps the parametric input (e.g., Kn) to the full velocity field. A physics-guided zonal loss is employed to improve prediction accuracy in the most
non-equilibrium regions of the flow.
The computational domain is partitioned into a recirculation zone $\Omega_r$,
identified by the reverse-flow criterion $u(x,y)<0$, and a background region $\Omega_b$.
The total training loss is defined as
\begin{equation}
\mathcal{L}
=
\lambda_r \mathcal{L}_r
+
\lambda_b \mathcal{L}_b,
\qquad \lambda_r > \lambda_b,
\end{equation}
where $\mathcal{L}_r$ and $\mathcal{L}_b$ denote mean-squared errors evaluated over
$\Omega_r$ and $\Omega_b$, respectively.
This targeted weighting prevents deceptively low global errors and significantly
improves fidelity in the separation bubble and shear layer, where strong gradients
and rarefaction effects dominate.

\begin{figure*}[t]
\centering

\begin{subfigure}[t]{\columnwidth}
\centering
\includegraphics[width=0.95\linewidth]{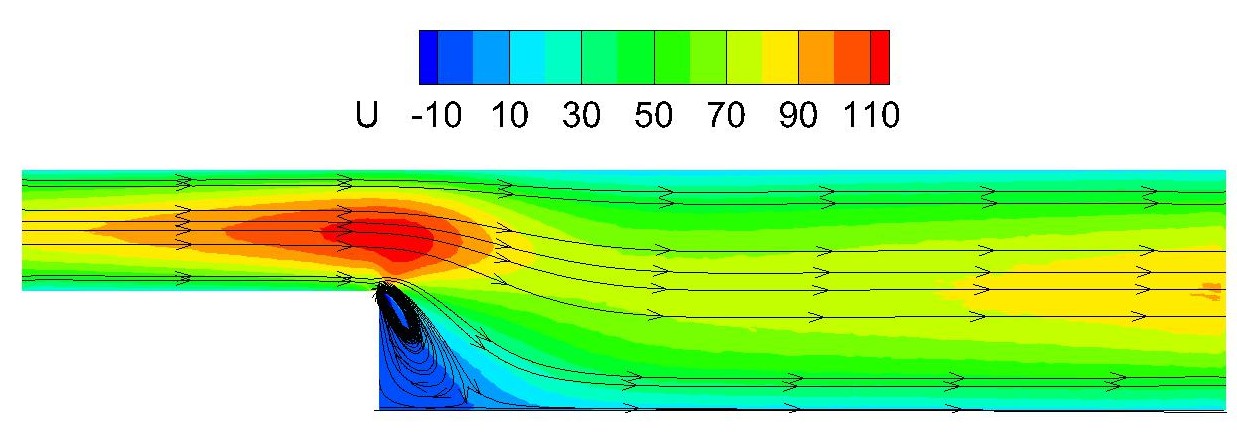}
\subcaption{U-velocity (DSMC ground truth)}
\label{fig:dsmc_U_kn002}
\end{subfigure}%
\hspace*{\columnsep}%
\begin{subfigure}[t]{\columnwidth}
\centering
\includegraphics[width=0.95\linewidth]{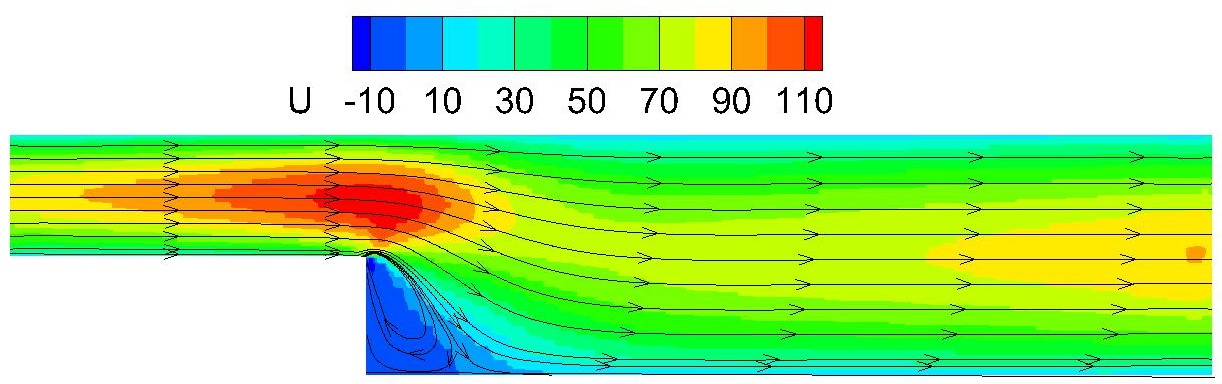}
\subcaption{U-velocity (DeepONet prediction)}
\label{fig:nn_U_kn002}
\end{subfigure}

\vspace{0.35cm}

\begin{subfigure}[t]{\columnwidth}
\centering
\includegraphics[width=0.95\linewidth]{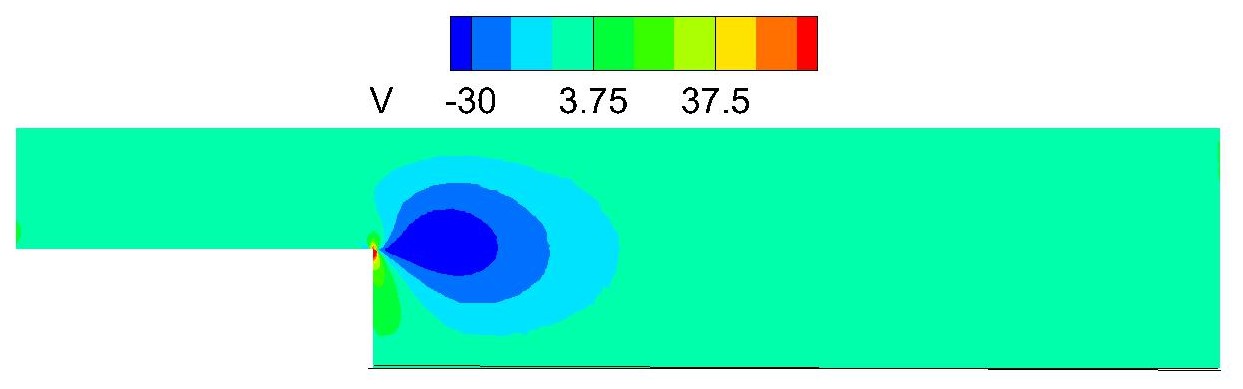}
\subcaption{V-velocity (DSMC ground truth)}
\label{fig:dsmc_V_kn002}
\end{subfigure}%
\hspace*{\columnsep}%
\begin{subfigure}[t]{\columnwidth}
\centering
\includegraphics[width=0.95\linewidth]{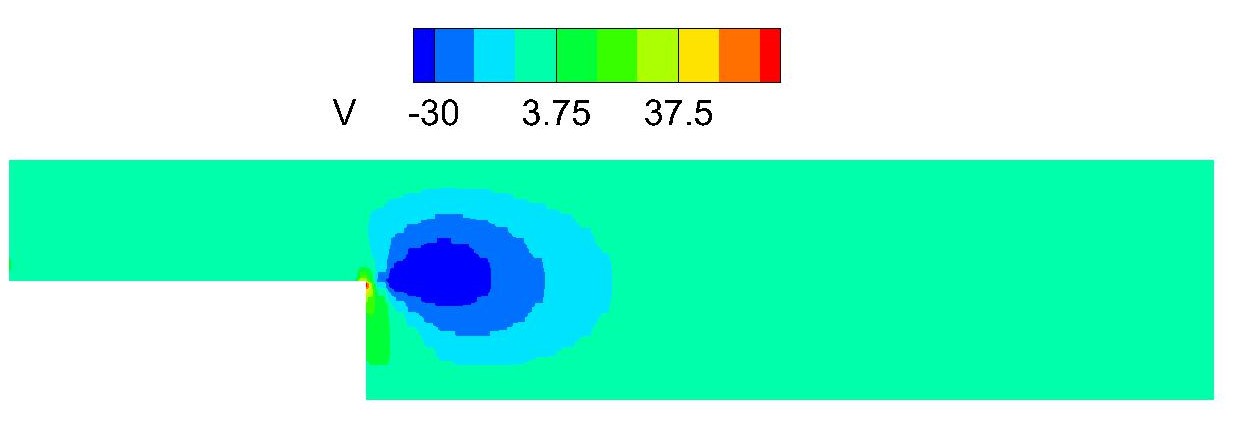}
\subcaption{V-velocity (DeepONet prediction)}
\label{fig:nn_V_kn002}
\end{subfigure}

\caption{
\textbf{Qualitative comparison of velocity fields at $\mathrm{Kn}=0.02$ (unseen test case).}
Comparison of streamwise ($U$) and wall-normal ($V$) velocity contours obtained from
high-fidelity DSMC simulations (ground truth) and DeepONet predictions.
The neural operator accurately reproduces the separated flow structure, recirculation
bubble, and shear-layer development despite this Knudsen number not being included
in the training dataset.
}
\label{fig:contour_comparison_UV_002}

\end{figure*}

Qualitative validation is demonstrated by direct comparisons of DSMC and DeepONet-predicted velocity contours for representative rarefaction regimes (e.g., Kn$=0.004$, 0.02, and 1), showing that the surrogate preserves the main flow topology and captures both the dominant streamwise velocity and the more delicate cross-stream component. In particular, Fig.~\ref{fig:contour_comparison_UV_002} highlights the Kn$=0.02$ test case,
which is not included in the training set.
The DeepONet prediction accurately reproduces the size and location of the recirculation
bubble, the shear-layer structure, and the overall velocity-field topology in both the
streamwise and wall-normal components when compared against the DSMC reference.

\begin{figure*}[t]
\centering

\begin{subfigure}[t]{\columnwidth}
\centering
\includegraphics[width=0.95\linewidth]{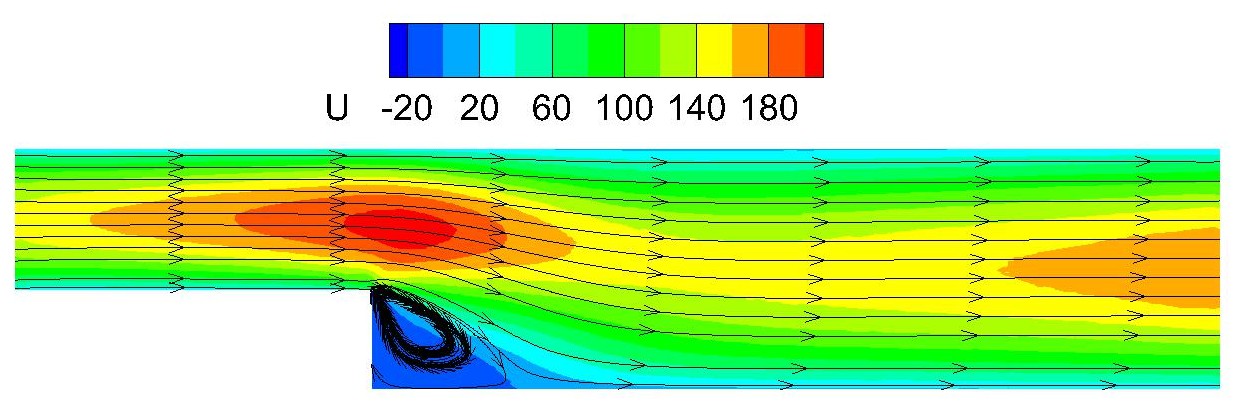}
\subcaption{U-velocity (DSMC ground truth)}
\label{fig:dsmc_U_h44}
\end{subfigure}%
\hspace*{\columnsep}%
\begin{subfigure}[t]{\columnwidth}
\centering
\includegraphics[width=0.95\linewidth]{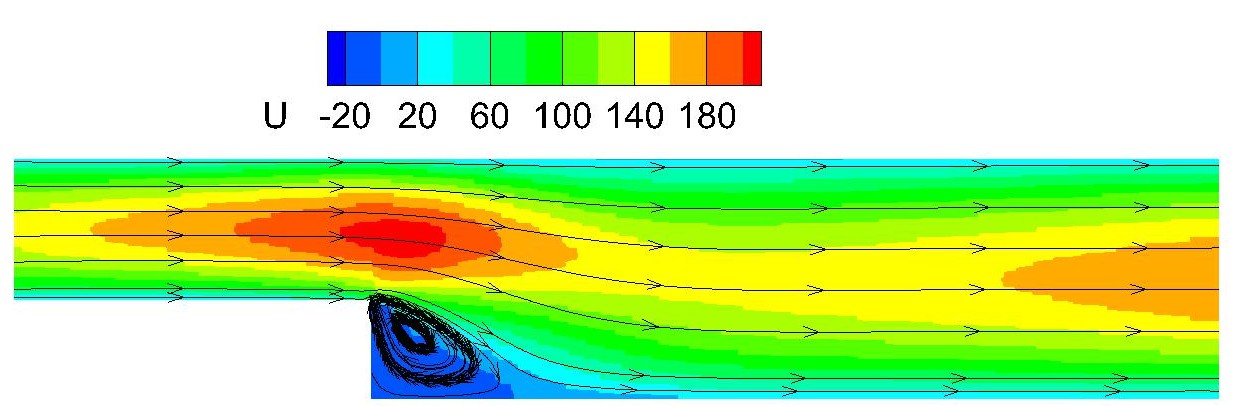}
\subcaption{U-velocity (DeepONet prediction)}
\label{fig:nn_U_h44}
\end{subfigure}

\vspace{0.35cm}

\begin{subfigure}[t]{\columnwidth}
\centering
\includegraphics[width=0.95\linewidth]{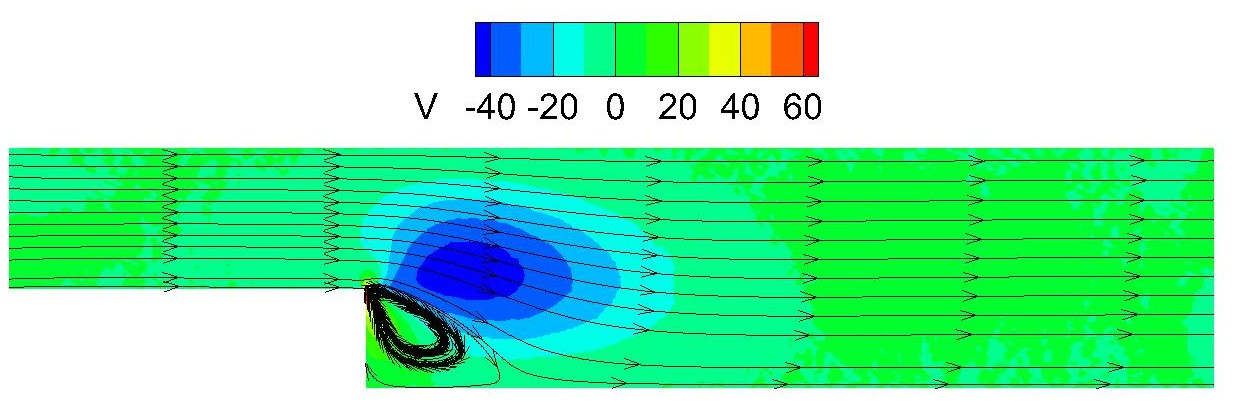}
\subcaption{V-velocity (DSMC ground truth)}
\label{fig:dsmc_V_h44}
\end{subfigure}%
\hspace*{\columnsep}%
\begin{subfigure}[t]{\columnwidth}
\centering
\includegraphics[width=0.95\linewidth]{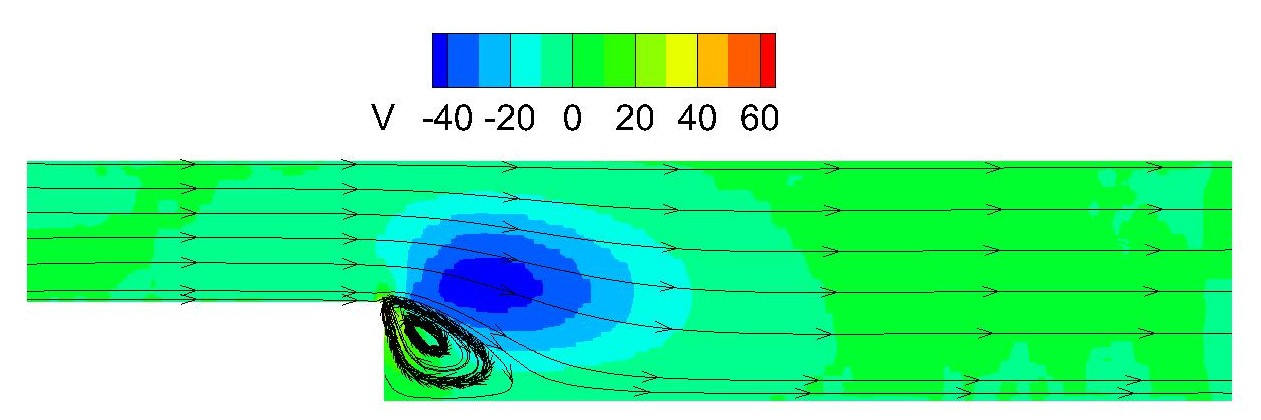}
\subcaption{V-velocity (DeepONet prediction)}
\label{fig:nn_V_h44}
\end{subfigure}

\caption{
\textbf{Generalization to an unseen geometry ($h/H=44\%$).}
Qualitative comparison of streamwise ($U$) and wall-normal ($V$) velocity contours
between the high-fidelity DSMC solution (ground truth) and the DeepONet surrogate.
Despite this step-height ratio not being included in the training dataset,
the neural operator accurately captures the separation bubble, shear-layer structure,
and overall flow topology.
}
\label{fig:height_comparison_h44}

\end{figure*}

Finally, the framework is extended to geometric parameterization by varying the step-height ratio \(h/H\) at fixed Kn$=0.01$, demonstrating the ability of the neural operator to learn geometry-to-solution mappings relevant to shape optimization. DSMC snapshots across multiple \(h/H\) values reveal a monotonic increase in separation extent and downstream shift of the recirculation center as \(h/H\) increases. In a data-scarce setting, two step-height ratios (\(h/H=44\%\) and \(67\%\)) are held out for testing; nevertheless, the DeepONet surrogate maintains strong qualitative agreement with DSMC for these unseen geometries, correctly identifying the recirculation zone and overall channel flow structure, see Fig.~\ref{fig:height_comparison_h44} for \(h/H=44\%\).


\subsection{Micro-Nozzle Flow: Generalization across pressure ratio and Knudsen number}

The fusion-DeepONet framework is further assessed on its ability to generalize across
operating conditions that strongly influence rarefied micro-nozzle flows, including
variations in pressure ratio and Knudsen number.
These parameters govern shock location, expansion strength, and the degree of
thermodynamic non-equilibrium, posing a challenging test for data-driven surrogate models.

\begin{figure}[htbp]
    \centering

    \begin{subfigure}[b]{0.49\textwidth}
        \centering
        \includegraphics[width=\linewidth]{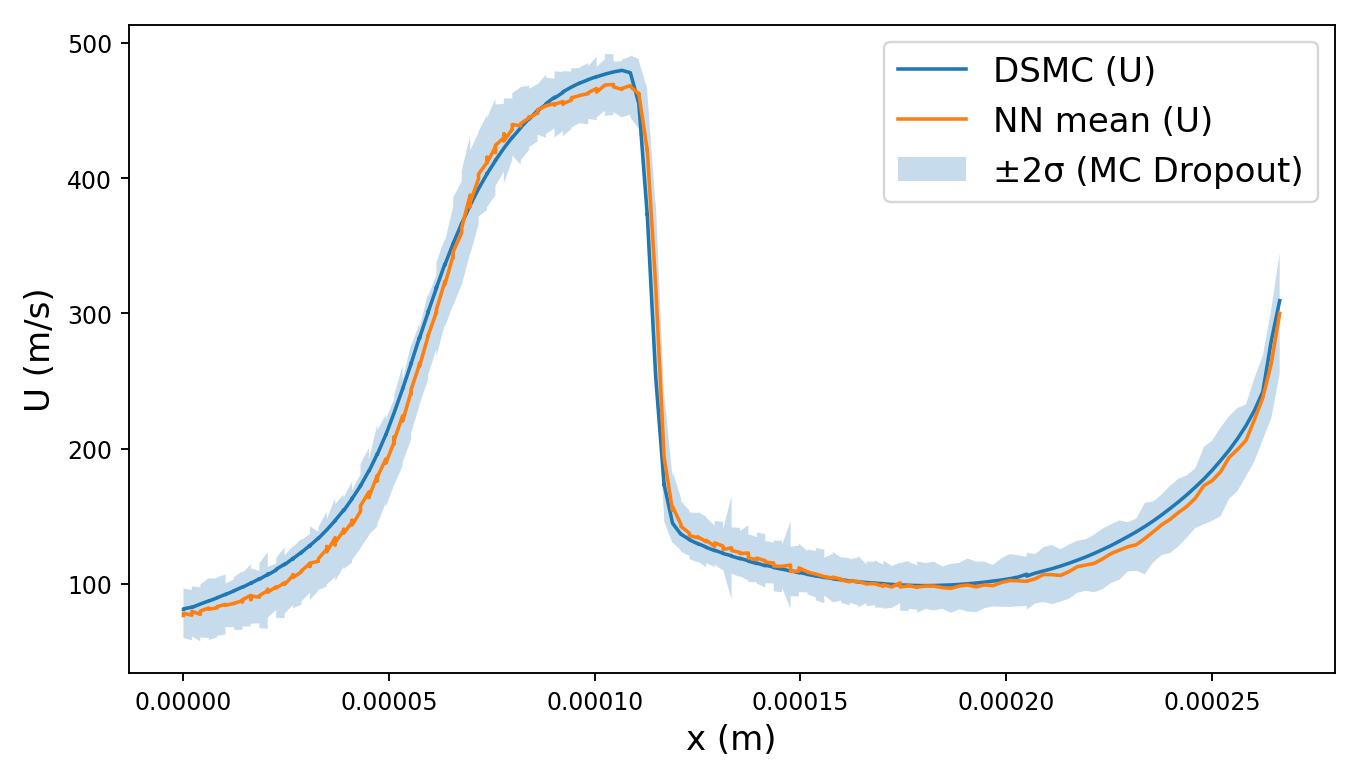}
        \subcaption{Back pressure = 25 kPa}
        \label{fig:centerline_U_pr25}
    \end{subfigure}
    \hfill
    \begin{subfigure}[b]{0.49\textwidth}
        \centering
        \includegraphics[width=\linewidth]{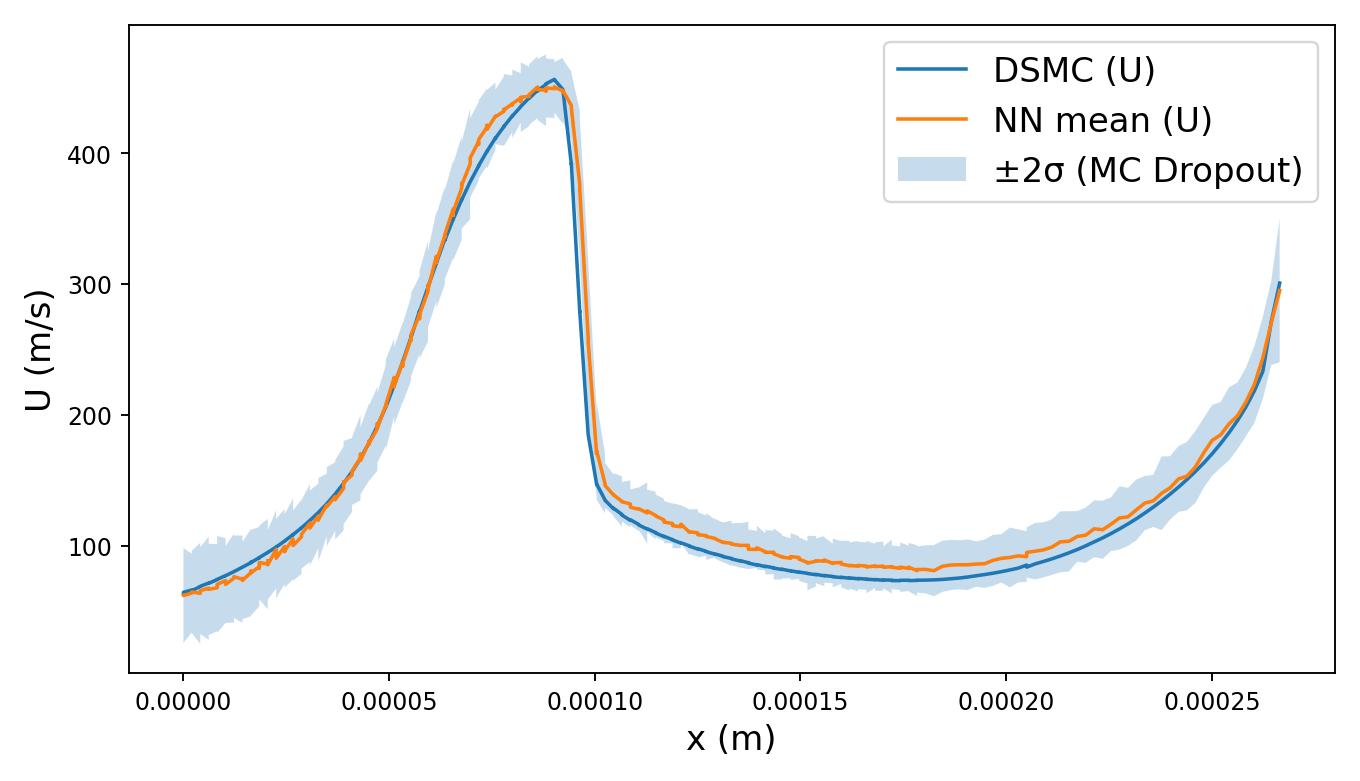}
        \subcaption{Back pressure = 30 kPa}
        \label{fig:centerline_U_pr30}
    \end{subfigure}

    \caption{
    Centerline axial velocity comparison between DSMC reference data and the
    fusion-DeepONet surrogate for two back-pressure conditions.
    The model accurately captures the pressure-ratio-induced changes in acceleration
    and downstream relaxation.
    }
    \label{fig:centerline_U_comparison}
\end{figure}

Figure~\ref{fig:centerline_U_comparison} examines the effect of back pressure on the
centerline axial velocity for two representative operating conditions, corresponding
to back pressures of 25~kPa and 30~kPa.
The DSMC reference solutions exhibit clear differences in the acceleration and relaxation
of the flow along the nozzle centerline as the pressure ratio changes.
The fusion-DeepONet surrogate accurately reproduces these trends, capturing both the
peak velocity and the downstream decay without introducing spurious oscillations.
This agreement demonstrates that the learned operator remains robust under variations
in pressure ratio that alter the global flow structure.

\begin{figure}[t]
  \centering
  \begin{subfigure}[t]{0.39\textwidth}
    \centering
    \includegraphics[width=\linewidth]{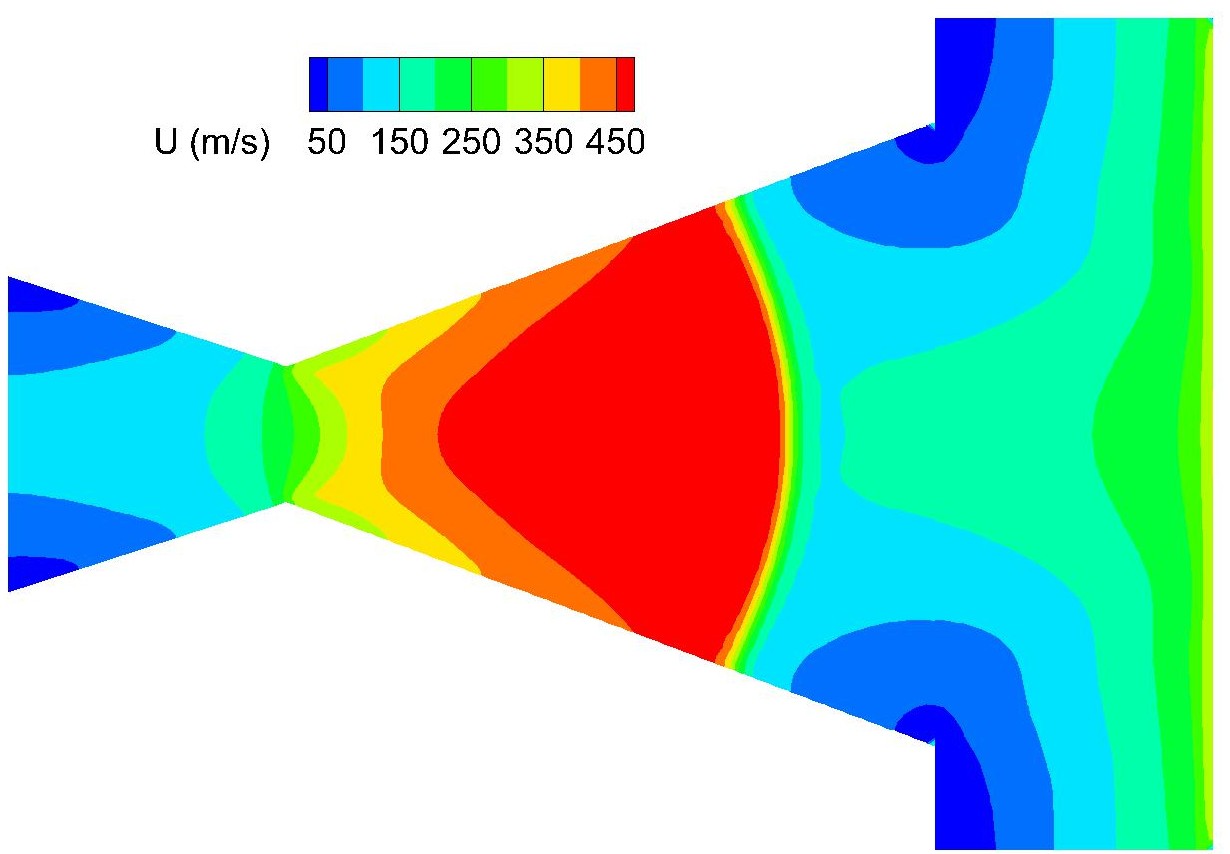}
    \subcaption{DSMC, $U$ at $X_{\mathrm{throat}}/L=0.30$}
    \label{fig:U_dsmc_30}
  \end{subfigure}
  \hfill
  \begin{subfigure}[t]{0.39\textwidth}
    \centering
    \includegraphics[width=\linewidth]{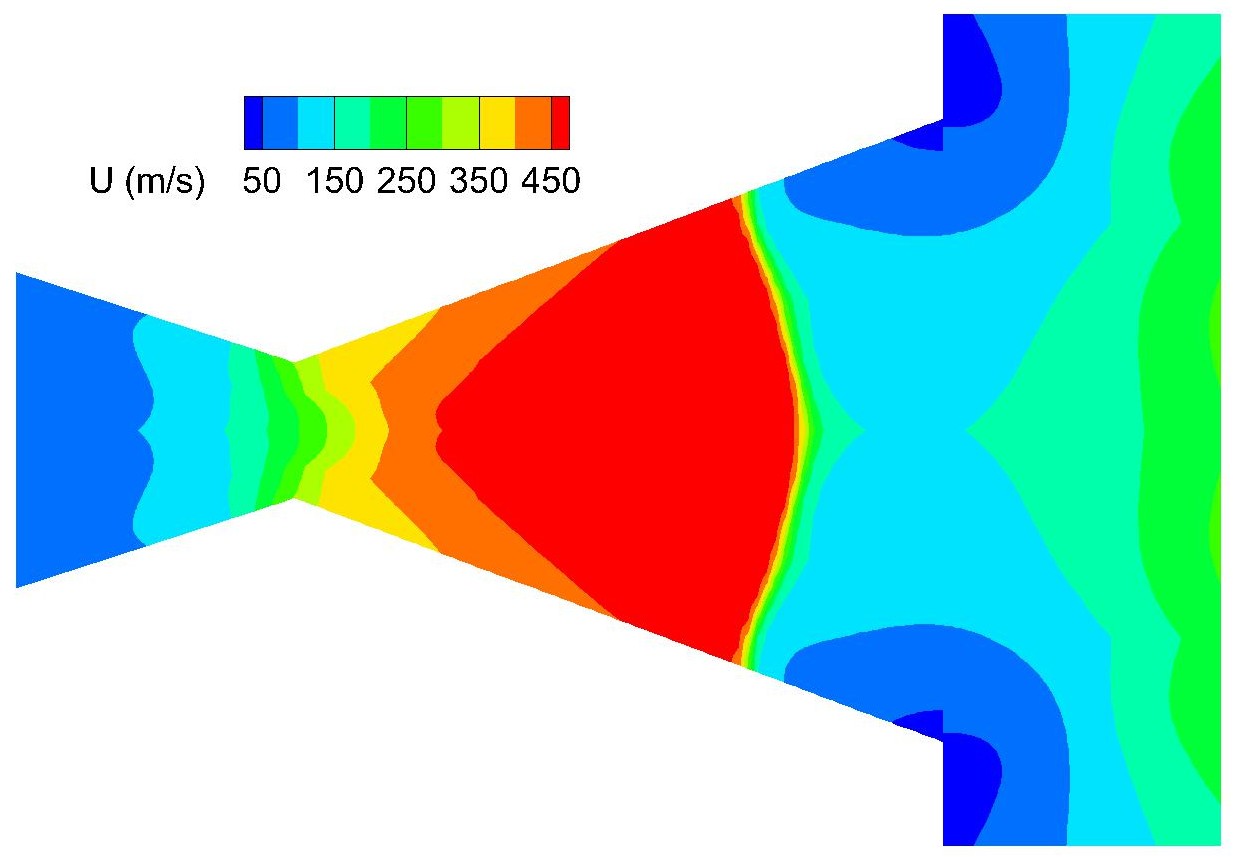}
    \subcaption{Fusion-DeepONet, $U$ at $X_{\mathrm{throat}}/L=0.30$}
    \label{fig:U_nn_30}
  \end{subfigure}

  \caption{
  Streamwise velocity contours at the nozzle throat location for a representative
  Knudsen-number test case.
  The fusion-DeepONet prediction closely matches the DSMC solution, preserving the
  spatial structure and magnitude of the velocity field under rarefied conditions.
  }
  \label{fig:U_compare_30}
\end{figure}

In addition to pressure effects, the influence of rarefaction is evaluated through
comparisons at different Knudsen numbers.
Figure~\ref{fig:U_compare_30} presents velocity-magnitude contours of the streamwise
component at a fixed throat location, highlighting the differences between the DSMC
solution and the neural-network prediction.
Despite the increased sensitivity of the flow field to molecular effects in this regime,
the surrogate successfully preserves the spatial distribution of velocity, including
regions of strong acceleration and shear.
The close correspondence between DSMC and fusion-DeepONet results confirms that the
proposed model can reliably generalize across Knudsen numbers, making it suitable for
many-query applications such as parametric sweeps and rapid design exploration.

\section{Conclusion}
This paper presented an AI-accelerated operator-learning framework to reduce the computational bottleneck of kinetic rarefied-flow solvers while preserving DSMC-level fidelity. A GPU-native DNN closure removed the dominant moment-closure cost in particle Fokker--Planck solvers and achieved near-Amdahl-limit speedups, while physics-guided and shock-aware DeepONet variants enabled data-efficient surrogate modeling across micro-step, micro-nozzle, and hypersonic cylinder cases. The proposed surrogates demonstrated strong generalization (including extrapolation to extreme conditions) and provided uncertainty estimates where applicable, supporting reliable many-query tasks such as parametric sweeps and design exploration. Overall, these results establish a practical pathway toward real-time, physics-consistent surrogate modeling for rarefied gas dynamics.

\section*{Acknowledgment}

The author would like to acknowledge Dr.~Ahmad Shoja-Sani and Dr.~Amirmehran Mahdavi
for their valuable contribution on providing DSMC results used in this paper for training of the neural networks.


\nocite{*}

\bibliographystyle{asmeconf}  
\bibliography{asmeconf-sample}

\appendix
\section{Network Architectures and Training Hyperparameters}

This appendix summarizes the neural-network architectures, training settings,
and datasets used across the different test cases presented in this work.
The intent is to improve reproducibility and clarify the relationship between
the various surrogate models employed.

\begin{table}[h!]
\centering
\caption{DNN hyperparameters for GPU-native Fokker--Planck closure surrogate.}
\label{tab:fp_dnn}
\begin{tabular}{ll}
\hline
Component & Setting \\
\hline
Network type & Fully connected MLP \\
Input features & 16 low-order velocity moments \\
Output dimension & 9 closure coefficients \\
Hidden layers & 3 \\
Neurons per layer & 64 \\
Activation function & ReLU \\
Optimizer & Adam \\
Learning rate & $10^{-3}$ \\
Training samples & $\sim 10^5$ local cell states \\
Loss function & Mean squared error (MSE) \\
Deployment & GPU-native (CuPy) \\
\hline
\end{tabular}
\end{table}

\begin{table}[h!]
\centering
\caption{DeepONet configuration for rarefied backward-facing step flows.}
\label{tab:step_deeponet}
\begin{tabular}{ll}
\hline
Component & Setting \\
\hline
Operator input & Knudsen number, geometry ratio $h/H$ \\
Output field & 2D velocity field $(u,v)$ \\
Branch network & 4 layers, 64 neurons/layer \\
Trunk network & 4 layers, 64 neurons/layer \\
Activation & Tanh \\
Loss function & Physics-guided zonal loss \\
Zonal criterion & $u(x,y)<0$ (recirculation) \\
Loss weights & $\lambda_r > \lambda_b$ \\
Optimizer & Adam \\
Learning rate & $10^{-4}$ \\
Training Kn range & $0.004 \leq \mathrm{Kn} \leq 1$ \\
Unseen tests & Kn = 0.02; $h/H = 44\%, 67\%$ \\
\hline
\end{tabular}
\end{table}

\begin{table}[h!]
\centering
\caption{Fusion-DeepONet hyperparameters for micro-nozzle surrogate modeling.}
\label{tab:nozzle_fusion}
\begin{tabular}{ll}
\hline
Component & Setting \\
\hline
Operator inputs & Pressure ratio, Knudsen number \\
Output field & 2D velocity field \\
Feature fusion & Geometry + physics-informed embedding \\
Branch network & 5 layers, 128 neurons/layer \\
Trunk network & 4 layers, 64 neurons/layer \\
Activation & Swish \\
Optimizer & Adam \\
Learning rate & $5\times10^{-4}$ \\
Uncertainty method & MC Dropout ($\pm2\sigma$) \\
Training cases & Multiple pressure ratios \\
Test cases & Unseen back pressures, Kn values \\
\hline
\end{tabular}
\end{table}

\vspace{-2pt}

\begin{table}[h!]
\centering
\caption{DeepONet ensemble configuration for hypersonic cylinder flow.}
\label{tab:cylinder_ensemble}
\begin{tabular}{ll}
\hline
Component & Setting \\
\hline
Operator input & Freestream Mach number \\
Mach range (train) & $5 \leq M \leq 14$ \\
Mach test (extrap.) & $M = 15$ \\
Output fields & Mach, pressure, temperature \\
Branch/trunk depth & 4 layers each \\
Neurons per layer & 64 \\
Activation & Tanh \\
Ensemble size & 5 networks \\
Uncertainty metric & Ensemble variance ($\pm2\sigma$) \\
Optimizer & Adam \\
Learning rate & $10^{-4}$ \\
\hline
\end{tabular}
\end{table}

\end{document}